# White light emission from silicon nanoparticles

Chengyun Zhang,[1,2,§] Yi Xu,[3,§] Jin Liu,[4] Juntao Li,[4] Jin Xiang,[1] Hui Li,[1] Jinxiang Li,[1] Qiaofeng Dai,[1] Sheng Lan[1*] and Andrey E. Miroshnichenko[5*]

[1]*Guangdong Provincial Key Laboratory of Nanophotonic Functional Materials and Devices, School of Information and Optoelectronic Science and Engineering, South China Normal University, Guangzhou 510006, P. R. China*

[2]*School of Physics and Electronic Engineering, Guangzhou University, Guangzhou 510006, China*

[3]*Department of Electronic Engineering, College of Information Science and Technology, Jinan University, Guangzhou 510632, China*

[4]*State Key Laboratory of Optoelectronic Materials and Technologies, Sun Yat-Sen University, Guangzhou, 510275, China*

[5]*Nonlinear Physics Centre, Research School of Science and Engineering, Australian National University, Canberra ACT 0200, Australia*

[*]*Corresponding authors:* slan@scnu.edu.cn, andrey.miroshnichenko@anu.edu.au; [§]Authors contributed equally to this work.

As one of the most important semiconductors, silicon (Si) has been used to fabricate electronic devices, waveguides, detectors, and solar cells etc. However, its indirect bandgap hinders the use of Si for making good emitters. For integrated photonic circuits, Si-based emitters with sizes in the range of 100–300 nm are highly desirable. Here, we show that efficient white light emission can be realized in spherical and cylindrical Si nanoparticles with feature sizes of ~200 nm. The up-converted luminescence appears at the magnetic and electric multipole resonances when the nanoparticles are resonantly excited at their magnetic and electric dipole resonances by using femtosecond (fs) laser pulses with ultralow low energy of ~40 pJ. The lifetime of the white light is as short as ~52 ps, almost three orders of magnitude smaller than the state-of-the-art results reported so far for Si (~10 ns). Our finding paves the way for realizing efficient Si-based emitters compatible with current semiconductor fabrication technology, which can be integrated to photonic circuits.

Owing to the existence of distinct electric and magnetic dipole (ED and MD) resonances in the visible to near infrared spectral range, high-index dielectric nanoparticles such as Si nanospheres (NSs) and nanopillars (NPs) with feature sizes ranging from 100 to 300 nm are considered as promising building blocks for the metamaterials operating at optical frequencies where artificial metallic atoms fail to work[1-7]. So far, the linear optical properties of Si nanoparticles have been extensively studied for different functionalities[2, 3, 6-20]. Very recently, much attention has been paid to the nonlinear optical responses of Si NPs by exploiting their MD resonances[21-28].

It is generally thought that Si is not suitable for making good emitters. Although Si quantum dots with sizes smaller than 10 nm have shown efficient visible light emission[29], they are difficult to be integrated with other devices. Apart from the quantum confinement effect[29, 30], luminescence from Si can be achieved by the radiative recombination of hot carriers provided that the radiative recombination lifetime can be significantly reduced to be comparable with the relaxation time (0.1−1.0 ps). According to the Fermi golden rule, the spontaneous emission rate of hot carriers can be increased by enhancing the electric field inside Si nanoparticles. In practice, the enhancement of the electric field has been realized by using plasmonic effect and hot luminescence has been demonstrated in Si nanowires (NWs) coated with thin silver films[5]. In this letter, we show that efficient two- and three-photon-induced luminescence (2PL and 3PL) can be achieved in Si nanoparticles by exploiting the significantly enhanced electric field at their electromagnetic multipole resonances.

We fabricated monocrystalline Si NSs by using fs laser ablation of a silicon-on-insulator (SOI) wafer and regularly arranged Si NPs by using electron beam lithography in combination with reactive ion etching (see Methods). The crystalline phase of the Si NSs was characterized by transmission electron microscope (TEM) (see Supplementary Information, Fig. S1). In Fig. 1a, we present the scattering spectrum of a Si NS with a diameter (*d*) of 192 nm calculated by Mie theory. It can be decomposed into the contributions of MD, ED, magnetic quadrupole (MQ) and electric quadrupole (EQ) resonances (see Supplementary Information, Fig. S2). A schematic showing the excitation and emission of the up-converted luminescence of a Si NS is shown in the inset of Fig. 1a. Basically, the 2PL intensity can be described as follows[31, 32]:

$$I_{2PL} = \eta(\lambda_{em})\delta(\lambda_{ex})|E_0|^4 L^4(\lambda_{ex})L^2(\lambda_{em})$$
$$\propto \eta(\lambda_{em})\delta(\lambda_{ex})|E_0|^4 (\frac{1}{V}\int_{Si}|E(\lambda_{ex},r)/E_0|^4 dV)(\frac{1}{V}\int_{Si}|E(\lambda_{em},r)/E_0|^2 dV) \quad (1)$$

where $\eta$ is the quantum efficiency, $\delta$ is the cross section of two-photon-induced absorption (2PA), $L(\lambda_{ex}) = |E(\lambda_{ex},r)/E_0|$ and $L(\lambda_{em}) = |E(\lambda_{em},r)/E_0|$ are the electric field enhancement factors at the excitation ($\lambda_{ex}$) and emission ($\lambda_{em}$) wavelengths, respectively. By replacing $\frac{1}{V}\int_{Si}|E(\lambda_{ex},r)/E_0|^4 dV$ with $\frac{1}{V}\int_{Si}|E(\lambda_{ex},r)/E_0|^6 dV$ in equation (1), one can obtain the formula describing the 3PL intensity qualitatively. In Fig. 1b, we present the spectra of $I^3 = \frac{1}{V}\int_{Si}|E(\lambda,r)/E_0|^6 dV$, $I^2 = \frac{1}{V}\int_{Si}|E(\lambda,r)/E_0|^4 dV$ and $I = \frac{1}{V}\int_{Si}|E(\lambda,r)/E_0|^2 dV$ calculated for the Si NS. The electric and magnetic field distributions at the first four resonances are also provided. It can be seen that the enhancement factor for $I^2$ (or $I^3$) is more than 50 (or 500) at the MD resonance of the Si NS. In addition, the enhancement factor for $I$ is ~5 at the MQ resonance. It implies that an enhancement in 2PL (or 3PL) of more than two (or three) orders of magnitude can be achieved if one excites the Si NS at the MD resonance and detects the 2PL (or 3PL) at the MQ resonance. Although the energy of one photon at the MD resonance (~758 nm) is not efficient for inducing a direct transition to the conduction band at the Γ point, the large enhancement in $|E|^4$ (or $|E|^6$) at the MD resonance would result in a significant enhancement in the 2PA (or 3PA), generating a large number of hot carriers. The large carrier density in combination with the electric field enhancement at the MQ and EQ resonances would lead to a dramatic reduction in the radiative recombination time $\tau_r$, making it comparable to the relaxation time. As a result, up-converted luminescence originating from the recombination of hot carriers would be observed, as schematically shown in Fig. 1c. If the energy of two photons is slightly smaller than the bandgap energy at the Γ point, the population of the conduction band through 2PA can be assisted by Rabi oscillation or phonons[5]. In addition, the 3PA process may become dominant at high excitation intensities.

In Fig. 2a, we show the scattering spectrum measured for a Si NS with $d$ ~192 nm whose scanning electron microscopy (SEM) image is shown in the inset. We varied the excitation wavelength ($\lambda_{ex}$) and examined the nonlinear optical response of the Si NS. For $\lambda_{ex}$ = 800 nm, we observed strong second harmonic generation (SHG) at 400 nm and very weak up-converted luminescence peaking at ~556 nm, as shown in Fig. 2b. Since the energy of two photons (~3.1 eV) is smaller than the bandgap energy at the Γ point (~3.4 eV), the hot carries are generated mainly by 3PA which is not significant at this excitation wavelength. When $\lambda_{ex}$ was tuned to 775 nm, a significant increase of the up-converted luminescence was observed at the MQ and EQ resonances and the nonlinear response spectrum became dominated by the up-converted luminescence (see Fig. 2c). For resonant excitation ($\lambda_{ex}$ = 758 nm), the up-converted luminescence was further enhanced by nearly one order of magnitude and the SHG was significantly suppressed (see Fig. 2d). In this case, efficient white light emission can be achieved by using pulse energy as small as ~40 pJ, as shown in the inset of Fig. 2d. These results demonstrate clearly that the up-converted luminescence of a Si NS can be significantly enhanced by resonantly exciting its MD resonance. We also examined the nonlinear optical responses of Si NSs with different diameters by using $\lambda_{ex}$ = 800 nm. It was found that the intensity of the SHG is reduced while that of the up-converted luminescence is increased rapidly when the MD resonance approaches the excitation wavelength (see Supplementary Information, Fig. S3).

In Fig. 3a, we present the nonlinear response spectra of another Si NS with *d* ~210 nm measured at different pulse energies (see Supplementary Information, Fig. S4). The spectrum of *I* is also provided for comparison. It is found that the peaks in the two spectra almost coincide, verifying the enhanced radiative recombination rate achieved at the MQ and EQ resonances. From the dependence of the up-converted luminescence on the excitation pulse energy plotted in the inset, a slope close to 2.9 was extracted, indicating that the up-converted luminescence is dominated by 3PL. In order to confirm the enhanced radiative recombination rate, we measured the luminescence lifetime at 590 nm, as shown in Fig. 3b. A lifetime as short as ~52 ps was extracted from the exponential decay of the luminescence (see also Supplementary Information, Fig. S5). This value is almost three orders of magnitude smaller than the shortest lifetime reported for Si nanostructures and it explains the physical origin for the observation of strong hot luminescence[5, 33, 34].

Although arrays of regularly arranged Si NSs have been successfully fabricated by using fs laser ablation[11], it is highly desirable that Si nanoparticles emitting strong visible light can be achieved by using the state-of-the-art fabrication technologies of semiconductor nanostructures[4]. Therefore, Si NPs are considered as the most suitable alternative and various functional devices composed of Si NPs have been experimentally demonstrated[3, 8, 16-19, 21-27]. In Fig. 4a, we show the scattering spectra measured and calculated for a single Si NP in a regular array of Si NPs fabricated on a quartz substrate (see Methods for details). The diameter, height and period of the Si NP were estimated to be ~148 nm, ~220 nm and ~5 μm, respectively. It can be seen that the MD (~720 nm) and ED (~620 nm) resonances dominate the scattering process of the Si NP (see Supplementary Information, Figs. S6). The nonlinear response spectra of the Si NP resonantly excited at 720 nm are shown in Fig. 4b. Similar to case of Si NSs, the resonant excitation of the MD resonance of the Si NP can also generate very efficient white light emission. Apart from the SHG appearing at 360 nm, it is noticed that the peaks of the up-converted luminescence are in good agreement with those predicted by the numerical simulation (see Fig. 4a). We also observed efficient up-converted luminescence from Si NPs fabricated directly on a SOI wafer with a technique compatible with the current fabrication technology of Si chips[4] (see Supplementary Information, Figs. S7, S8 and S9) .

Although the up-converted luminescence observed in Si NSs and NPs seems similar to that reported previously in Si NWs coated with Ag films[5], the mechanism for boosting up the electric field inside Si is substantially different. In our case, the electric field enhancement is mediated by the electromagnetic multipole resonances of Si nanoparticles rather than the lossy plasmonic effect. More importantly, pure Si nanoparticles can be easily integrated into current Si-based photonic circuits. Further improvement in up-converted luminescence can be expected by engineering the electromagnetic modal properties of Si nanoparticles.

In summary, we have proposed and demonstrated a novel strategy for white light emission from Si nanoparticles by exploiting their electromagnetic resonances. It was shown by numerical simulation that the up-converted luminescence of a Si nanoparticle can be enhanced by more than three orders of magnitude through resonantly exciting the MD resonance of the Si nanoparticle. In experiments, efficient while light emission was demonstrated in Si NSs and the luminescence lifetime as short as ~52 ps was observed. Furthermore, we have also shown that efficient up-converted luminescence can also be generated in Si NPs, which are compatible with the current fabrication technology of Si-based chips. Our finding might be a solid step towards the achievement of on-chip light sources for silicon photonics[4].

**Acknowledgements**

The authors would like to thank the comments and suggestions from Prof. Dragomir N. Neshev and Prof. Yuri S. Kivshar. We also acknowledge the financial support from the National Nature and Science Foundation of China (Grant Nos. 11374109, 11674110 and 11674130), the Natural Science Foundation of Guangdong Province, China (Grant Nos. 2016A030308010, 2015A030310382 and 2016A030306016), the National Key Research Program of China (No. 1016YFA0201002), the Science and Technology Planning Project of Guangdong Province, China (Grant No. 2015B090927006), the Science and Technology Program of Guangzhou (No. 201607010261).

**Author contributions**

S. L., A. E. M., Y. X., and C. Y. Z. conceived the idea. C. Y. Z., J. X. and Y. X. fabricated the Si NPs and carried out the optical experiments. J. L. and J. T. Li. fabricated the Si NPs. Y. X. and J. X. performed the numerical modeling. S. L., Y. X., A. E. M. and C. Y. Z. analyzed the data and wrote the manuscript. S. L. and A. E. M. supervised the project. All the authors read and commented on the manuscript.

**Figures and captions:**

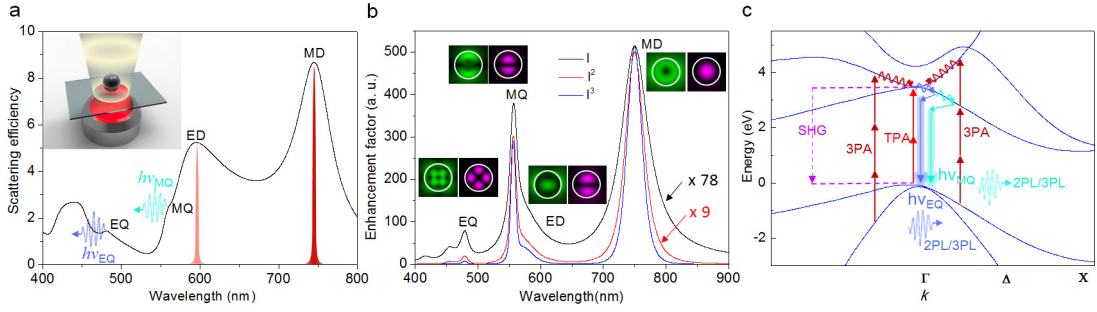

**Fig. 1** (a) Principle of utilizing the MD and ED resonances of a Si NS to realize the excitation enhancement in 2PA and 3PA and the MQ and EQ resonances to realize the emission enhancement in 2PL and 3PL. The inset shows schematically the excitation of a Si NS placed on a quartz substrate by using fs laser light in the near inferred spectral range (red color) and the emission of up-converted luminescence (white light). The red pulse-like symbols indicate the excitation wavelengths. (b) Spectra of $I$, $I^2$ and $I^3$ calculated for a Si NS with a diameter $d = 192$ nm. The electric (left) and magnetic (right) intensity distributions calculated at the MD, ED, MQ and EQ resonances are presented as insets. (c) Energy band diagram of Si in which the carrier excitation process through 2PA/3PA and the photon emission processes through SHG and 2PL/3PL are schematically depicted. The enhanced radiative recombination at the MQ and EQ resonances is highlighted by color arrows.

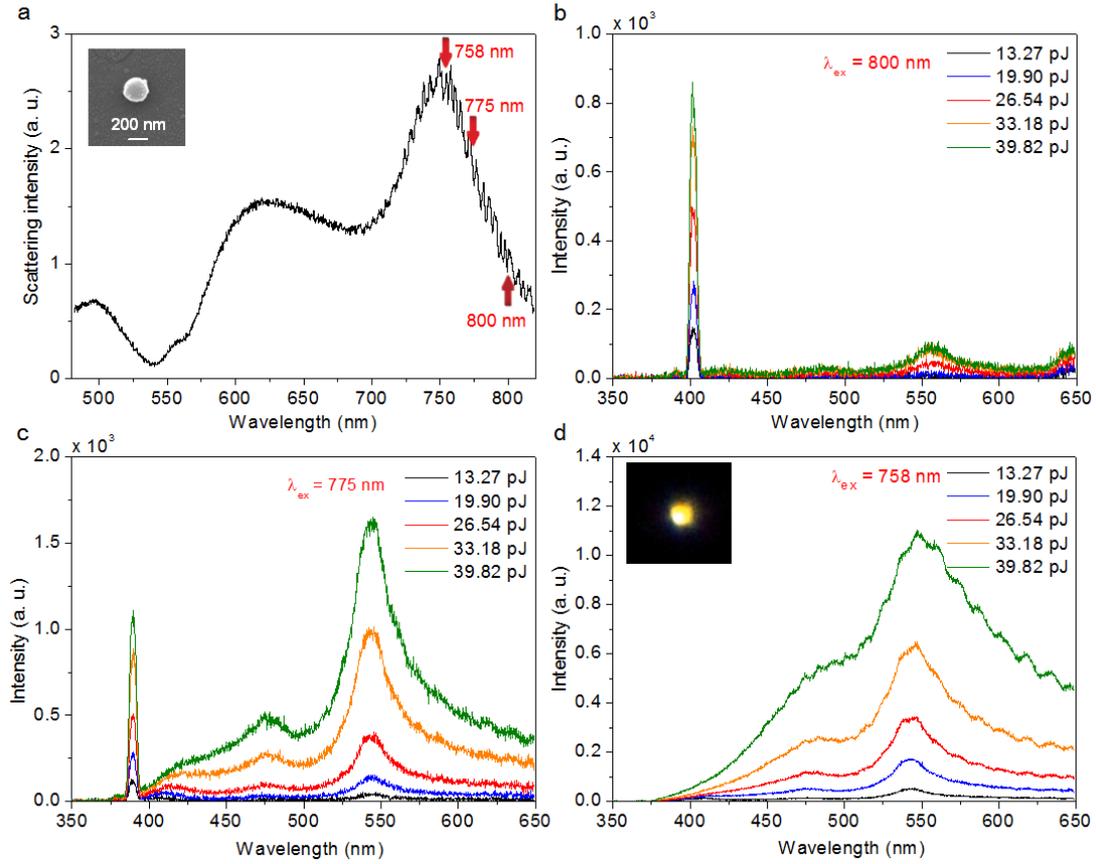

**Fig. 2** (a) Scattering spectra measured for a Si NS with $d \sim 192$ nm. The inset shows the SEM image of the Si NS. The red arrows indicate the excitation wavelengths of the fs laser in the experiments. (b), (c) and (d) show the dependence of the nonlinear response spectrum of the Si NS on the excitation pulse energy measured at different excitation wavelengths of 800, 775, and 758 nm, respectively. White light emission recorded by a charge coupled device is shown in the inset of (d).

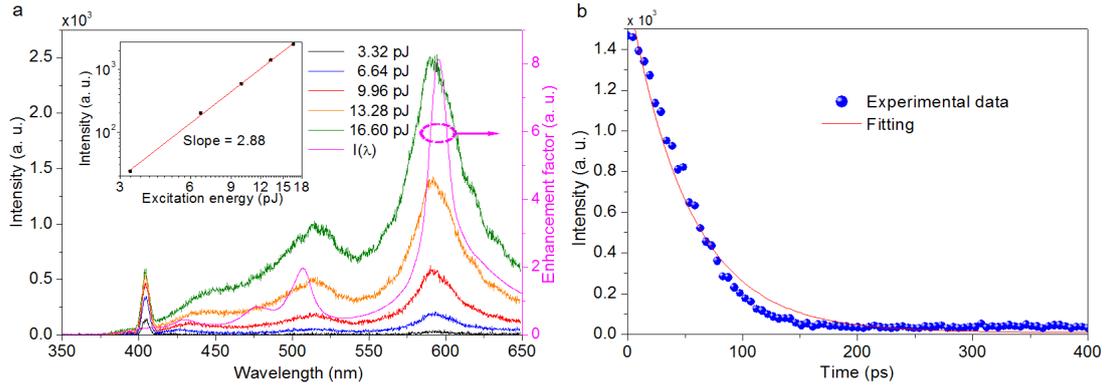

**Fig. 3** (a) Nonlinear response spectra measured at different excitation pulse energies for a Si NS with $d \sim 210$ nm. The excitation wavelength is 804 nm. The spectrum of $I$ calculated for the Si NS (i.e. enhancement of electric field intensity inside the Si NS) is also presented for comparison. The inset shows the dependence of the up-converted luminescence on the excitation pulse energy plotted in a double-logarithmic coordinate. (b) Decay of the up-converted luminescence after the excitation of the fs laser pulses (16.60 pJ) measured for the Si NS at 590 nm. The luminescence intensity is fitted by an exponential decay of $6.76+1683.08e^{-t/52}$.

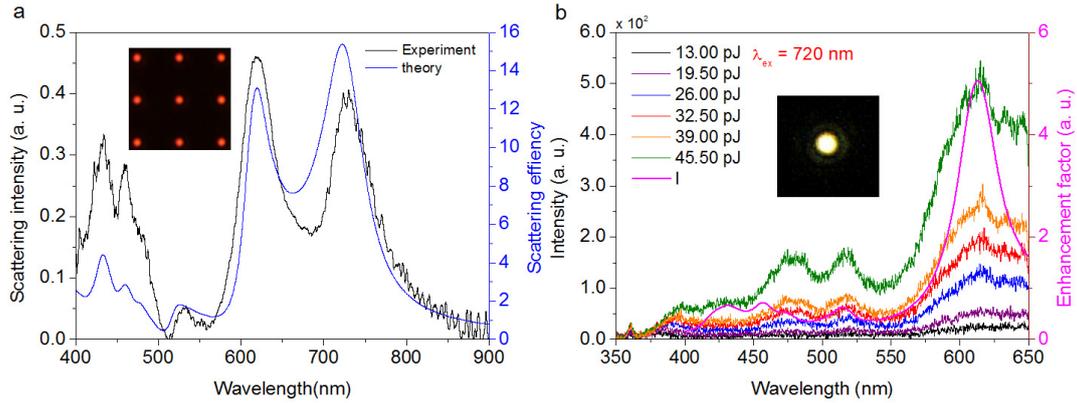

**Fig. 4** (a) Scattering spectra calculated and measured for a single Si NP in a regular array of Si NPs on a quartz substrate whose dark-field microscope image is shown in the inset. The diameter, height and period of the Si NP were estimated to be ~148 nm, ~220 nm and ~5 μm, respectively. (b) Nonlinear response spectra measured for the Si NP resonantly excited at its MD resonance (~720 nm) with different pulse energies. Spectrum of *I* calculated for the Si NP is also provided. The inset shows the charge coupled device image of the up-converted luminescence.

# Supplementary Information

## Table of contents

1.  **Transmission electron microscope image of the Si nanoparticle fabricated by fs laser ablation**

The transmission electron microscope (TEM) image of a Si nanosphere (NS) fabricated by using fs laser ablation is shown in Fig. S1. The crystalline phase of the Si NS is indicated by paralleled red lines. There is a very thin amorphous shell on the surface of the Si NS.

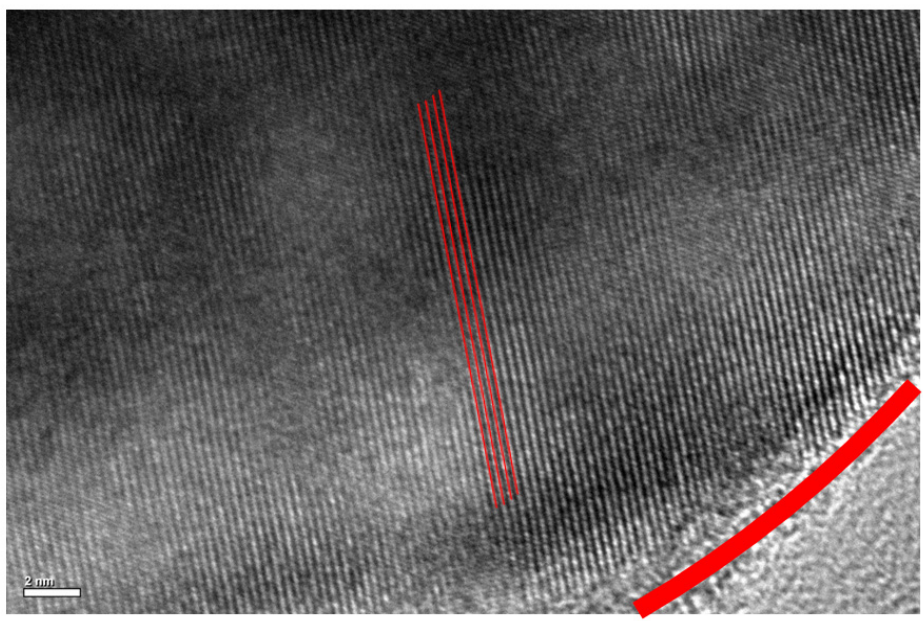

**Fig. S1** TEM image of a monocrystalline Si NS. The crystalline phase of the Si NS is indicated by paralleled red lines and the outer surface is indicated by a bold red curve. The length of the scale bar is 2 nm.

2.  **Analysis of the electric and magnetic multipole resonances based on Mie theory**

Based on Mie theory, the total scattering of a nanoparticle can be decomposed into the contributions of electric and magnetic resonances of different orders. As an example, we show the decomposition of the scattering spectrum of a Si NS with $d$ = 192 nm into the contributions of MD, ED, MQ, and ED resonances. It is noticed that except the ED resonance, all the resonances exhibit large quality factors. For example, quality factors of ~38 and ~27 are observed for the MQ and EQ resonances, implying that a large enhancement of the electric field can be achieved at these resonances. This feature indicates the possibility of realizing efficient excitation at the MD resonance and achieving efficient up-converted luminescence emission at the high-order electromagnetic resonances at the same time, such as the MQ and EQ resonances.

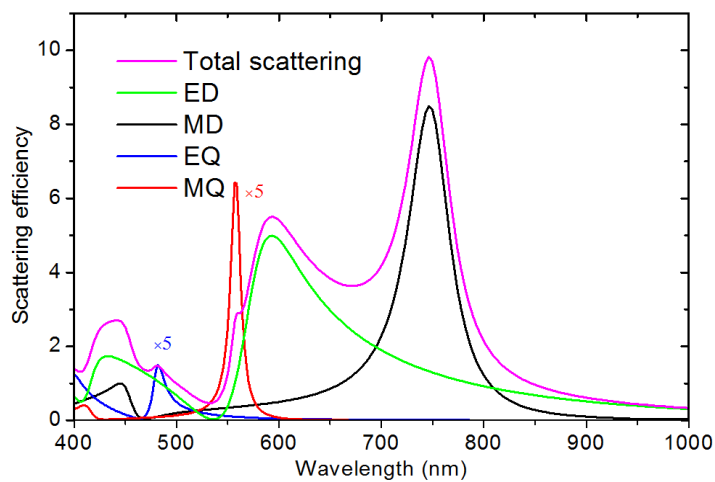

**Fig. S2** Scattering spectrum of a Si NS with $d$ = 192 nm calculated based on Mie theory. The contributions of ED, MD, EQ and MQ resonances to the total scattering are also presented.

3. **Nonlinear response spectra and up-converted luminescence lifetime of Si nanoparticles**

We have examined the nonlinear optical responses of Si NSs with different diameters of $d$ = 103, 122 and 160 nm. The results measured at different pulse energies by using $\lambda_{ex}$ = 800 nm are shown in Fig. S3. The images of the scattering light taken by a charge coupled device are presented in the insets, which appears to be blue, green and red respectively. In each case, it is noticed that the scattering spectrum is dominated by the MD resonance which is shifted to longer wavelengths with increasing the diameter of the Si NS. For the Si NS with $d$ = 103 nm, only SHG was observed in the nonlinear response spectrum, as shown in Fig. S3b. In comparison, very weak up-converted luminescence was found at the long-wavelength side of the second harmonic for the Si NS with $d$ = 122 nm, as shown in Fig. S3d. For the Si NS with $d$ = 160 nm, it was observed that the intensity of the SHG decreased while that of the up-converted luminescence increased and became stronger than the SHG intensity, as shown in Fig. S3f. These results indicate clearly that the MD resonance of a Si NS plays crucial in enhancing the up-converted luminescence.

In Fig. S4, we present the linear and nonlinear optical properties of another Si NS with a larger diameter of $d$ = 210 nm. Similar to the Si NS whose linear and nonlinear properties are shown in Fig. 2, it is noticed that the up-converted luminescence of the Si NS is also maximized when the MD resonance is resonantly excited. In order to further confirm the dramatic reduction in the luminescence lifetime, we also measured the luminescence lifetime for a Si NS with a diameter of $d$ = 205 nm, as shown in Fig. S5. A lifetime as short as ~58 ps, which is similar to the one presented in Fig. 3b, was extracted from the exponential decay of the luminescence.

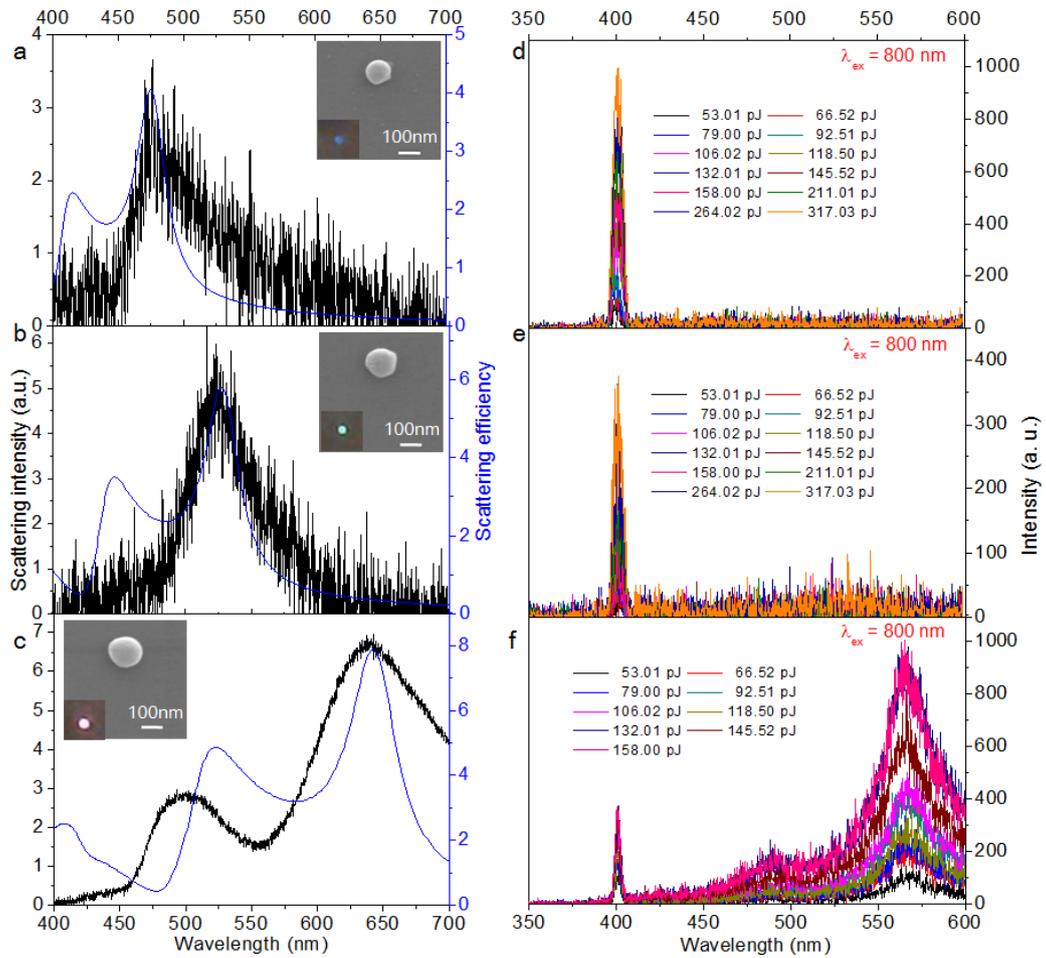

**Fig. S3** Scattering spectra calculated and measured for a Si NS with $d$ = 103 nm (a), $d$ = 122 nm (b) and $d$ = 160 nm (c). The SEM images and dark-field microscope images measured for the Si NSs are shown in the insets. The nonlinear response spectra for the Si NSs measured at different pulse energies are presented in (d), (e) and (f), respectively. The excitation wavelength was fixed at 800 nm.

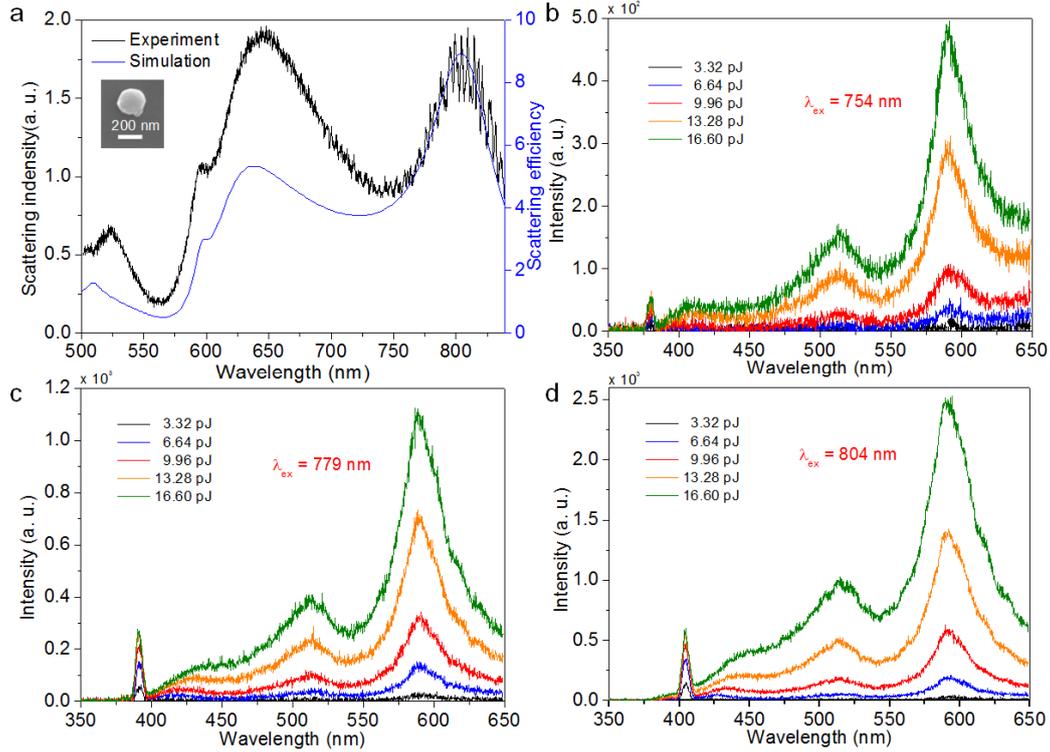

**Fig. S4** (a) Calculated and measured scattering spectra for a Si NS with $d$ = 210 nm. The SEM image for the Si NS is shown in the inset. The nonlinear response spectra of the Si NP measured at different excitation wavelengths of 754, 779, and 804 nm are shown in (b), (c), and (d), respectively.

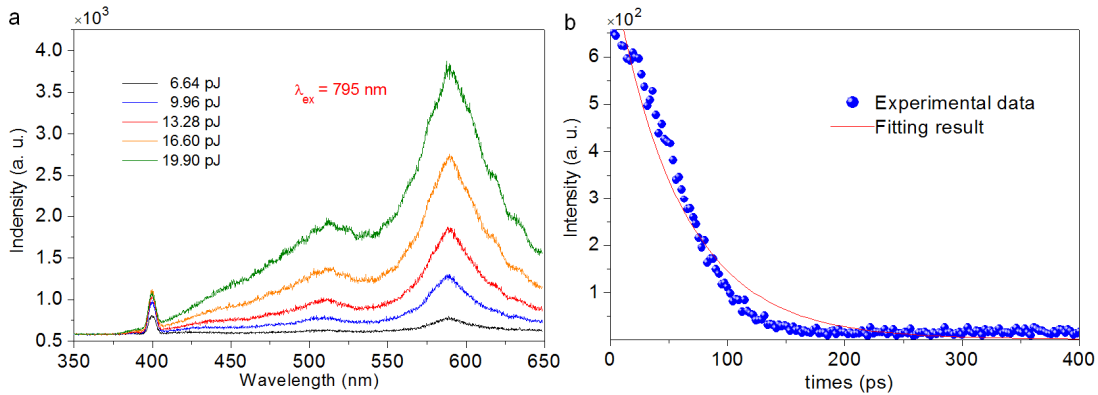

**Fig. S5** (a) Nonlinear response spectra measured for a Si NS with $d$ = 205 nm at different pulse energies. The excitation wavelength was chosen to be 795 nm. (b) Decay of the up-converted luminescence, which was fitted by $1.87+793.52e^{-t/58}$, after the excitation of the fs laser pulses (19.90 pJ) measured at ~590 nm.

### 4. Multipolar decomposition based on spherical harmonics for a Si nanopillar

By utilizing spherical harmonics, the total scattering of a Si nanopillars (NP) can be decomposed into the contributions of electric and magnetic resonances of different orders. In Fig. S6, we show the decomposition results (up to quadrupole) for the scattering spectrum of a Si NP with $d = 148$ nm and $h = 220$ nm. It is noticed that MD and ED resonances dominate the scattering process.

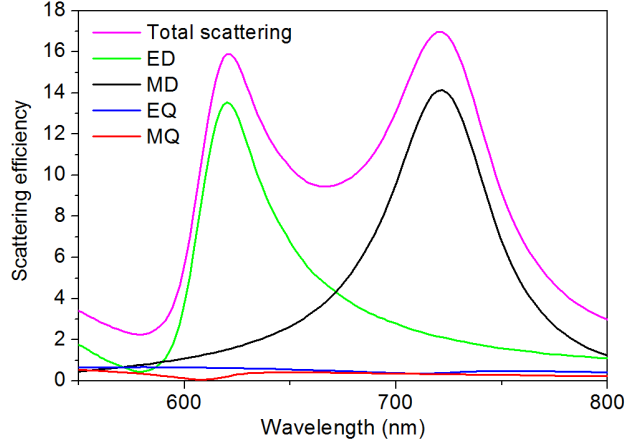

**Fig. S6** Scattering spectrum of a Si NP with $d = 148$ nm and $h = 220$ nm calculated by FDTD simulation. The contributions of ED, MD, EQ and MQ resonances to the total scattering are evaluated numerically utilizing spherical harmonics.

5. **Nonlinear response spectra of Si nanopillars fabricated on a silicon-on-insulator wafer**

The scanning electron microscope (SEM) images of the Si NPs fabricated by using electron beam lithography in combination with reactive ion etching are shown in Figs. S7 and S8. In Fig. S9a, we show the scattering spectra calculated for two Si NPs. One Si NP has a diameter of $d$ ~170 nm and the other is actually a truncated nanocone (NC) with diameters of $d_1 = 210$ and $d_2 = 250$ nm (see Figs. S7 and S8). It can be seen that the MD resonance of the Si NP appears at ~790 nm while the ED resonance of the truncated Si NC appears at ~760 nm. In Fig. S9b, we also present the spectra of $I^3$ calculated for them and enhancement factors quite similar to those of the Si NSs are observed, implying that efficient up-converted luminescence can also be achieved in Si NPs if $\lambda_{ex}$ is chosen at the MD or ED resonance. The nonlinear response spectra of the Si NP resonantly excited at 790 nm are shown in Fig. S9c. Apart from the SHG appearing at 395 nm, it is noticed that the up-converted luminescence peaks are in good agreement with those predicted by numerical simulation. We also performed confocal microscopic imaging of the 3PL for the array of Si NPs by using $\lambda_{ex} = 790$ nm and observed efficient up-converted luminescence from most Si NPs in the array, as shown in the inset of Fig. S9c. One can find some Si NPs in the array which didn't emit luminescence. It may be caused by the relatively large deviation of their sizes from the designed value, indicating the importance of resonant excitation in the generation of efficient up-converted luminescence. In Fig. S9d, we show the nonlinear response spectrum for the truncated Si NC measured at different pulse energies and its corresponding confocal microscopic image. Differently, we chose to resonantly excite the ED resonance of the Si NC by using $\lambda_{ex}$ ~760 nm. Similar to the excitation of the MD resonance, the resonant excitation of the ED resonance of the Si NC can also generate very efficient white light emission.

Since the substrate is not transparent for the Si NPs fabricated directly on a SOI wafer, we cannot measure the scattering spectra of such Si NPs by using our dark-field microscope. For this reason, the ED and MD resonances of the Si NPs were determined only by numerical simulation according to the sizes of the Si NPs estimated based on their SEM images. In addition, it is difficult to accurately locate a single Si NP without the help of bright field image in the measurements of up-converted luminescence, leading to a low excitation and collection efficiency. Consequently, the up-converted luminescence observed for the Si NPs fabricated directly on a SOI wafer is not as strong as that observed for the Si NPs fabricated on the transparent quartz substrate.

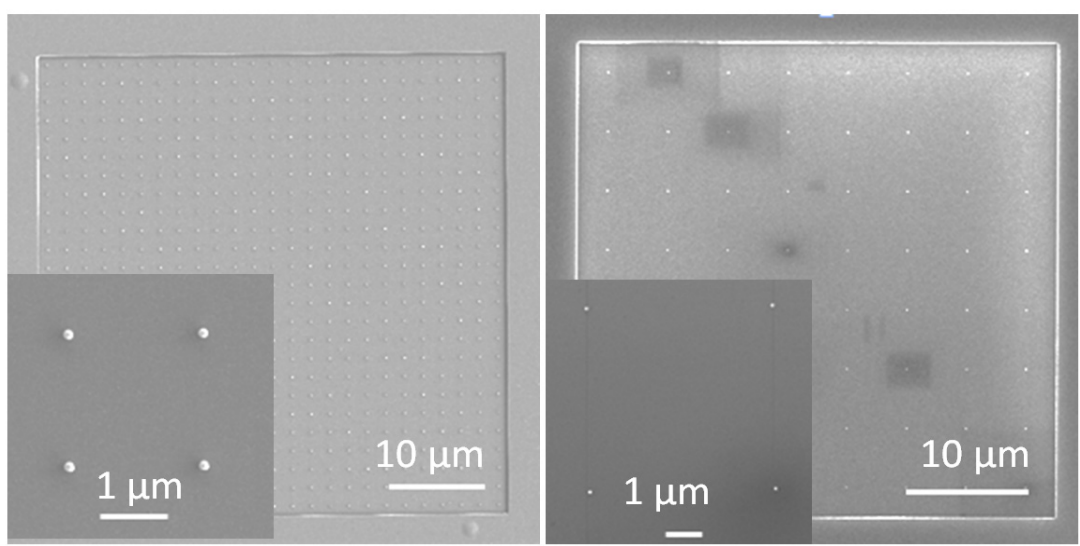

**Fig. S7** (a) SEM image for the array of Si NPs with $d$ = 170 nm, $h$ = 220 nm, and $p$ = 2 μm. (b) SEM image for the array of truncated Si NCs with $d_1$ = 210 nm, $d_2$ = 250 nm, $h$ = 220 nm, and $p$ = 5 μm. Here, $p$ is the period of the array.

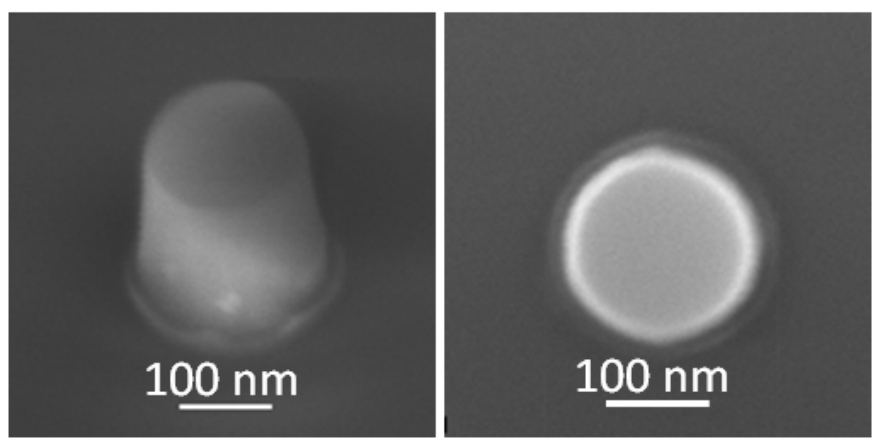

**Fig. S8** SEM images for a typical truncated Si NC used in the measurements of the nonlinear response spectra. (a) Side view. (b) Top view.

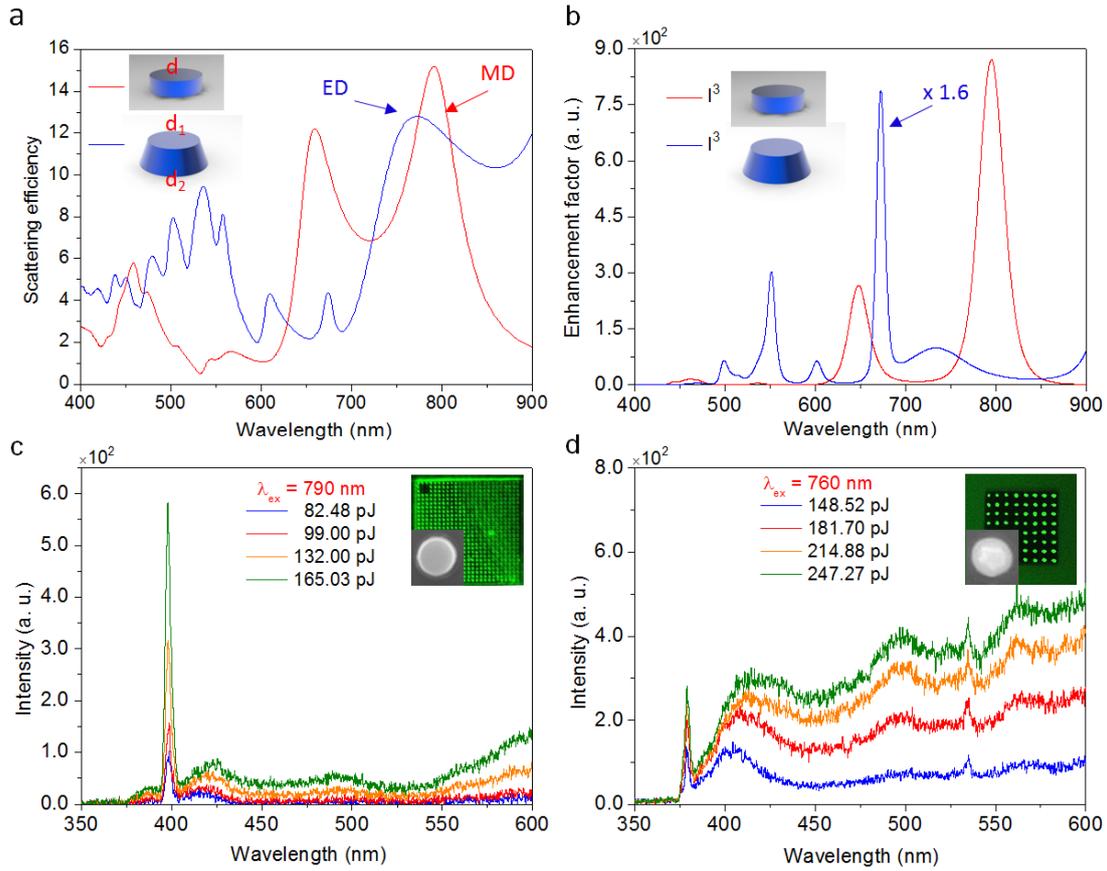

**Fig. S9** (a) Scattering spectra calculated for a Si NP with $d$ = 170 nm and a truncated Si NC with $d_1$ = 210 nm and $d_2$ = 250 nm. (b) Spectra of $I^3$ calculated for the Si NP and truncated Si NC. (c) Nonlinear response spectra measured with different pulse energies for a single Si NP which is excited at the MD resonance. (d) Nonlinear response spectra measured at different excitation pulse energies for a single truncated Si NC which is excited at the ED resonance. The insets show the emission patterns of the array of Si NPs obtained by using a confocal laser scanning microscope. In each case, the SEM image for a single Si NP or NC is also provided as an inset.